\documentstyle[preprint,tighten,pra,aps]{revtex}
\begin{document} \draft
\def\bbz{Z\!\!\!Z}
\def\bl{\lambda \kern-6.5pt \lambda}
\def\bH{ \bf H \rm }
\def\ch{{\cal H} }
\def\br{\rho \kern-5.5pt \rho}
\def\bbr{I\!\!R}
\def\bbc{C\kern-6.5pt I}
\def\bbn{I\!\!N}
\def\k{\kappa}
\def\o{\bar 0}
\def\I{\bar 1}
\def\pr{\perp}
\def\us{\parallel}
\def\ss{\subset}
\def\cD{{\cal D}}
\def\ci{{\cal I}}
\def\cf{{\cal F}}
\def\ct{{\cal T}}
\def\Uq{U_q(\cg)}
\def\a{\alpha} 
\def\b{\beta}
\def\d{\delta}
\def\ve{\varepsilon}
\def\e{\epsilon}
\def\vr{\vert}
\def\g{\gamma}
\def\G{\Gamma}
\def\x{\xi}
\def\p{\pi}
\def\y{\eta}
\def\l{\lambda}
\def\m{\mu}
\def\n{\nu}
\def\D{\Delta} 
\def\L{\Lambda}
\def\r{\rho}
\def\hr{{\hat\r}}
\def\hd{{\hat D}}
\def\om{\omega}
\def\s{\sigma}
\def\ca{{\cal A}}
\def\cb{{\cal B}}
\def\cg{{\cal G}}
\def\cd{{\cal D}}
\def\ch{{\cal H}}
\def\ck{{\cal K}}
\def\cl{{\cal L}}
\def\ce{{\cal E}}
\def\cm{{\cal M}}
\def\cn{{\cal N}}
\def\cf{{\cal F}}
\def\cs{{\cal S}}
\def\bk{ \bf {\cal K}}
\def\tn{{\tilde \cn}}
\def\cu{{\cal U}}
\def\cp{{\cal P}}
\def\cz{{\cal Z}}
\def\lra{\longleftrightarrow}
\def\ra{\longrightarrow}
\def\mt{\mapsto}
\def\h{\chi}
\def\th{\theta}
\def\Th{\Theta}
\def\t{\tau}
\def\O{\Omega}
\def\tl{\tilde \l}
\def\tk{\tilde k}
\def\tell{\tilde \ell}
\def\tm{\tilde m}
\def\tilh{\tilde h}
\def\td{\tilde D}
\def\sqr#1#2{{\vcenter{\vbox{\hrule height.#2pt
        \hbox{\vrule width.#2pt height#1pt \kern#1pt
           \vrule width.#2pt}
		\hrule height.#2pt}}}}
\def\square{\mathchoice\sqr68\sqr68\sqr{2.1}3\sqr{1.5}3}		   
\def\xtil{{\lower5pt \hbox{$\buildrel {\rm {\big x}}  \over \sim $ }} }
\def\ptil  {{\lower5pt \hbox{$\buildrel {\big p}  \over \sim $ }} }
\def\qtil {{\lower5pt \hbox{$\buildrel {\big q}  \over \sim $ }} }
\def\defeq {$\buildrel {\equiv} \over def $ }
\def\simeq {\buildrel {\sim} \over {=} }
\def\gl {\buildrel {>} \over {<}  }
\def\pnot{ ( p \cdot {p^'} )  }
\def\pdotp2 { { ( p \cdot {p^'}) }^2 }
\def\bX{\bf X }
\def\bY{\bf Y }
\def\bM{\bf M }
\def\bL{\bf L } 
\def\tbL{ \bf {\tilde L } } 
\def\tbA{ \bf {\tilde A  } }
\def\tbQ{ \bf {\tilde Q  } }
\def\tbD{ \bf { \tilde D } }
\def\tbC{ \bf { \tilde  C } }
\def\bC{\bf C }
\def\bG{\bf G }
\def\bw{\bf W }
\def\bT{\bf T }
\def\bP{\bf P }
\def\bH{\bf H }
\def\bW{\bf \omega }
\def\bQ{\bf Q }
\def\bD{\bf D }
\def\bSig{\bf \Sigma }
\def\cch{{\cal H}_{c}}
\def\ccg{{\cal G}_{c}}
\def\bGtil{\bf {\tilde G} }
\def\bGmtil{\bf {\tilde \G} }
\def\bKtil{\bf {\tilde K} }
\def\bGamtil{\bf {\tilde \Gamma} }
\def\ftil{\tilde f}
\def\halb{{1 \over 2}}
\def\vier{{1 \over 4}}
\def\5in2{{5 \over 2}}
\def\3in4{{3 \over 4}}
\def\threehalves{{\frac {3} {2}}}
\def\prm{ \prime}
\def\bNtil{\bf {\tilde N }}

\def\bS{ \bf{\cal S}}  
\def\Poin{\tilde {\bf {\cal P}} \rm }
\def\ddx0{ \matrix{ \partial \cr \overline {\partial x_{0}} \cr } }
\def\ddxj{ \matrix{ \partial \cr \overline {\partial x_{j}} \cr } }
\def\ddxe{ \matrix{ \partial \cr \overline {\partial x_{i}} \cr } }
\def\ddxio{ \matrix{ \partial \cr \overline {\partial \xi_{0}} \cr } }
\def\pzero{ \sqrt{ 1~+~ p^{2} } }
\def\ddxij{ \matrix{ \partial \cr \overline {\partial \xi_{j}} \cr } }
\def\ddxi4{ \matrix{ \partial \cr \overline {\partial \xi_{4}} \cr } }
\def\ddxii{ \matrix{ \partial \cr \overline {\partial \xi_{i}} \cr } }
\def\ddui{ \matrix{ \partial \cr \overline {\partial u_{i}} \cr } }
\def\dduj{ \matrix{ \partial \cr \overline {\partial u_{j}} \cr } }
\def\ddu4{ \matrix{ \partial \cr \overline {\partial u_{4}} \cr } }  
\def\ddpi{ \matrix{ \partial \cr \overline {\partial p_{i}} \cr } }  
\def\ddpj{ \matrix{ \partial \cr \overline {\partial p_{j}} \cr } }  
\def\ker{ \rm ker \it}
\def\Im{ \rm Im \it}
\def\imp{ \rm ~~~~implies ~~~~ \it}
\def\piro{ {\pi_\r} }
\def\sadj{ S^{*} }
\def\tadj{ T^{*} }
\def\tadjadj { T^{**}}
\def\aadj{ A^{*} }
\def\badj{ B^{*} }
\def\cadj{ C^{*} }
\def\dadj{ D^{*} }
\def\Uadj{ U^{*} }
\def\eq{ ~~=~~}
\def\defeq{ :=~~}
\def\adj{^{*}}
\def\tom{^{-1}}
\def\plus{~~+~~}
\def\minus{~~-~~}
\def\psubset{~~~~{\underline{\subset}}~~~~}
\def\ket{>}
\def\el{\ell}
\def\pitilde{ \tilde \pi}
\def\tworootq{[2]_{\sqrt q}}
\def\twoq{[2]_{\sqrt q}}

\def\NCA{\em Nuovo Cimento}
\def\NIM{\em Nucl. Instrum. Methods}
\def\NIMA{{\em Nucl. Instrum. Methods} A}
\def\NPB{{\em Nucl. Phys.} B}
\def\PLB{{\em Phys. Lett.}  B}
\def\PRL{\em Phys. Rev. Lett.}
\def\PRD{{\em Phys. Rev.} D}
\def\PRB{{\em Phys. Rev.} B}
\def\ZPC{{\em Z. Phys.} C}
\def\JMP{\em J. Math. Phys.}
\def\st{\scriptstyle}
\def\sst{\scriptscriptstyle}
\def\mco{\multicolumn}
\def\epp{\epsilon^{\prime}}
\def\vep{\varepsilon}
\def\ra{\rightarrow}
\def\ppg{\pi^+\pi^-\gamma}
\def\vp{{\bf p}}
\def\ko{K^0}
\def\kb{\bar{K^0}}
\def\al{\alpha}
\def\ab{\bar{\alpha}}
\def\tpi{ {\tilde {\pi} } }
\def\tbdel{\bf {\tilde {\Delta} } }
\def\bA{\bf A }
\def\tbA{ \bf {\tilde A  } }
\def\be{\begin{equation}}
\def\bdel{\bf {\Delta} }
\def\ee{\end{equation}}
\def\bea{\begin{eqnarray}}
\def\eea{\end{eqnarray}}
\def\CPbar{\hbox{{\rm CP}\hskip-1.80em{/}}}
\def\bbz{Z\!\!\!Z}
\def\bl{\lambda \kern-6.5pt \lambda}
\def\bH{ \bf H \rm }
\def\ch{{\cal H} }
\def\br{\rho \kern-5.5pt \rho}
\def\bbr{I\!\!R}
\def\bbc{C\kern-6.5pt I}
\def\bbn{I\!\!N}
\def\k{\kappa}
\def\o{\bar 0}
\def\I{\bar 1}
\def\pr{\perp}
\def\us{\parallel}
\def\ss{\subset}
\def\cD{{\cal D}}
\def\ci{{\cal I}}
\def\cf{{\cal F}}
\def\ct{{\cal T}}
\def\Uq{U_q(\cg)}
\def\a{\alpha} 
\def\b{\beta}
\def\d{\delta}
\def\ve{\varepsilon}
\def\e{\epsilon}
\def\vr{\vert}
\def\g{\gamma}
\def\G{\Gamma}
\def\x{\xi}
\def\p{\pi}
\def\y{\eta}
\def\l{\lambda}
\def\m{\mu}
\def\p{\pi}
\def\n{\nu}
\def\D{\Delta} 
\def\L{\Lambda}
\def\r{\rho}
\def\hr{{\hat\r}}
\def\hd{{\hat D}}
\def\om{\omega}
\def\s{\sigma}
\def\ca{{\cal A}}
\def\cb{{\cal B}}
\def\cg{{\cal G}}
\def\cd{{\cal D}}
\def\ch{{\cal H}}
\def\ck{{\cal K}}
\def\cl{{\cal L}}
\def\ce{{\cal E}}
\def\cm{{\cal M}}
\def\cn{{\cal N}}
\def\cf{{\cal F}}
\def\cs{{\cal S}}
\def\bk{ \bf {\cal K}}
\def\tn{{\tilde \cn}}
\def\cu{{\cal U}}
\def\cp{{\cal P}}
\def\cz{{\cal Z}}
\def\lra{\longleftrightarrow}
\def\ra{\longrightarrow}
\def\mt{\mapsto}
\def\h{\chi}
\def\th{\theta}
\def\Th{\Theta}
\def\t{\tau}
\def\O{\Omega}
\def\tl{\tilde \l}
\def\tk{\tilde k}
\def\tell{\tilde \ell}
\def\tm{\tilde m}
\def\tilh{\tilde h}
\def\td{\tilde D}
\def\sqr#1#2{{\vcenter{\vbox{\hrule height.#2pt
        \hbox{\vrule width.#2pt height#1pt \kern#1pt
           \vrule width.#2pt}
		\hrule height.#2pt}}}}
\def\square{\mathchoice\sqr68\sqr68\sqr{2.1}3\sqr{1.5}3}		   
\def\xtil{{\lower5pt \hbox{$\buildrel {\rm {\big x}}  \over \sim $ }} }
\def\ptil  {{\lower5pt \hbox{$\buildrel {\big p}  \over \sim $ }} }
\def\qtil {{\lower5pt \hbox{$\buildrel {\big q}  \over \sim $ }} }
\def\defeq {$\buildrel {\equiv} \over def $ }
\def\simeq {\buildrel {\sim} \over {=} }
\def\gl {\buildrel {>} \over {<}  }
\def\pnot{ ( p \cdot {p^'} )  }
\def\pdotp2 { { ( p \cdot {p^'}) }^2 }
\def\bX{\bf X }
\def\bY{\bf Y }
\def\bM{\bf M }
\def\bL{\bf L } 
\def\tbL{ \bf {\tilde L } } 
\def\tbA{ \bf {\tilde A  } }
\def\tbQ{ \bf {\tilde Q  } }
\def\tbD{ \bf { \tilde D } }
\def\tbC{ \bf { \tilde C } }
\def\bC{\bf C }
\def\bG{\bf G }
\def\bw{\bf W }
\def\bT{\bf T }
\def\bP{\bf P }
\def\bH{\bf H }
\def\bW{\bf \omega }
\def\bQ{\bf Q }
\def\bD{\bf D }
\def\bSig{\bf \Sigma }
\def\cch{{\cal H}_{c}}
\def\ccg{{\cal G}_{c}}
\def\bGtil{\bf {\tilde G} }
\def\bGmtil{\bf {\tilde \G} }
\def\bKtil{\bf {\tilde K} }
\def\bGamtil{\bf {\tilde \Gamma} }
\def\ftil{\tilde f}
\def\halb{{1 \over 2}}
\def\vier{{1 \over 4}}
\def\5in2{{5 \over 2}}
\def\3in4{{3 \over 4}}
\def\prm{ \prime}
\def\bNtil{\bf {\tilde N }}
\def\pirotilde{{\tilde \pi}}
\def\bS{ \bf{\cal S}}  
\def\Poin{\tilde {\bf {\cal P}} \rm }
\def\ddx0{ \matrix{ \partial \cr \overline {\partial x_{0}} \cr } }
\def\ddxj{ \matrix{ \partial \cr \overline {\partial x_{j}} \cr } }
\def\ddxe{ \matrix{ \partial \cr \overline {\partial x_{i}} \cr } }
\def\ddxio{ \matrix{ \partial \cr \overline {\partial \xi_{0}} \cr } }
\def\pzero{ \sqrt{ 1~+~ p^{2} } }
\def\ddxij{ \matrix{ \partial \cr \overline {\partial \xi_{j}} \cr } }
\def\ddxi4{ \matrix{ \partial \cr \overline {\partial \xi_{4}} \cr } }
\def\ddxii{ \matrix{ \partial \cr \overline {\partial \xi_{i}} \cr } }
\def\ddui{ \matrix{ \partial \cr \overline {\partial u_{i}} \cr } }
\def\dduj{ \matrix{ \partial \cr \overline {\partial u_{j}} \cr } }
\def\ddu4{ \matrix{ \partial \cr \overline {\partial u_{4}} \cr } }  
\def\ddpi{ \matrix{ \partial \cr \overline {\partial p_{i}} \cr } }  
\def\ddpj{ \matrix{ \partial \cr \overline {\partial p_{j}} \cr } }  
\def\ker{ \rm ker \it}
\def\Im{ \rm Im \it}
\def\imp{ \rm ~~~~implies ~~~~ \it}
\def\piro{ {\pi_\r} }
\def\sadj{ S^{*} }
\def\tadj{ T^{*} }
\def\tadjadj { T^{**}}
\def\aadj{ A^{*} }
\def\badj{ B^{*} }
\def\cadj{ C^{*} }
\def\dadj{ D^{*} }
\def\Uadj{ U^{*} }
\def\eq{ ~~=~~}
\def\defeq{ :=~~}
\def\adj{^{\dagger}}
\def\tom{^{-1}}
\def\plus{~~+~~}
\def\minus{~~-~~}
\def\psubset{~~~~{\underline{\subset}}~~~~}
\def\ket{>}
\def\el{\ell}
\def\pitilde{ \tilde \pi}
\def\tworootq{[2]_{\sqrt q}}
\def\twoq{[2]_{\sqrt q}}


\title{\LARGE \bf Tachyons and Representations of ${ \bf SO_0(2,3)}$}

\author{Patrick Moylan}
\address{Physics Department \\
The Pennsylvania State University \\
Abington College \\
Abington, Pennsylvania 19001 \ USA }

\maketitle

\begin{abstract} We have previously described 
an embedding 
of the Poincar\'e Lie algebra into an extension of the Lie field of the 
group $SO_0(1,4)$, and we used this embedding to construct irreducible 
representations of the Poincar\'e group out of representations of 
$SO_0(1,4)$.  Some $q$ generalizations of these results have been 
obtained by us i.e. we 
embed classical 
structures into quantum structures. Here  we report on analogous  
findings for $SO_0(2,3)$: we express the basis elements of the Lie algebra 
of the  Poincar\'e Lie group as irrational functions of the 
generators of the anti de Sitter group $SO_0(2,3)$, and thus obtain an 
embedding of the Poincar\'e Lie algebra into an extension of 
the Lie field of $SO_0(2,3)$. We have obtained some generalizations to higher 
dimensions. We apply our results
to certain unitary, continuous series representations of SO(2,3) associated with
functions on SO(2,3)/SO(1,3). From 
the embedding theorem we obtain representations of the Poincar\'e Lie 
algebra which are associated with unitary, tachyonic representations 
of the Poincar\'e Lie group, provided these representations are 
integrable to group representations.  Such tachyonic representations 
were studied by Wigner in his classical 1939 paper.

\end{abstract}

\vspace{8mm}

\section{Introduction}

We start with a well-known deformation \cite{1}, \cite{2}, \cite{3} 
of the Poincar\'e Lie algebra, which is defined in terms of the generators ${\bL}_{ij}$ (Lorentz 
generators) 
and ${\bP}_i$ (translation generators)   by the following:
$${\bL}_{ij} ~~\rightarrow ~~{\bL}_{ij}~~, \eqno(1a)$$
$${\bP}_i ~~\rightarrow ~~{\bL}^{\pm}_{4i} \eq  {{i~[ {\bQ}_2, ~{\bP}_i]}  \over 
2~\sqrt{\pm ~ \sum_{i,~j~=~0}^3 {\bP}_i ~
{\bP}^i}  
} 
~+~{\bP}_i ~~~\eqno(1b\pm)$$
where ${\bQ}~=~ \halb~ \sum_{i,~j~=~0}^3 ~{\bL}_{ij}~{\bL}^{ji}$ is the 
second order Casimir operator of the Lorentz subgroup. ($[~,~]$ denotes 
commutator.) 
For the plus sign this leads to the Lie algebra of the anti-de Sitter group,
 $SO_{0}(2,3)$, and the minus sign gives the commutation relations of the 
de Sitter group, $SO_0(1,4)$. Now eqns. (1b$\pm$) may be considered 
as algebraic 
equations for the translation generators  ${\bP}_i $ of the Poincar\'e 
group, and we are able to solve these equations for the ${\bP}_i $. The 
solution to this problem for the choice of the minus sign (i.e. for the 
choice of 
eqn. (1.b$-$) has been 
given by us in [1]. Our 
answer expresses the basis elements of the Poincar\'e Lie algebra as 
certain irrational functions of the generators of the  
de Sitter group. We give in section 3 a generalization 
of this ``anti-deformation'' to higher dimensions i.e. for $SO_0(p,q)$ 
groups and associated higher dimensional Poincar\'e groups, but only for 
a particular class of representations, which in the four dimensional
case considered in [1] corresponds to spinless representations of the 
Poincar\'e group. It is possible to obtain the general solution 
for eqn. (1.1+) also, and we present this solution in section 4.

We have also obtained some analogous results for q-deformations in 
lowest dimensions i.e. 2, 3 and 4.  In particular, in the $n=2$ case, 
we start with 
the Euclidean group in two dimensions ${\cal E}(2)$, with generators 
${\bf L}_{12}$ (rotation generator) and ${\bf P}_i$ ($i=1,2$) (translation 
generators), and define the following(c.f. [4]): ( $[m]_{q} ~=~  { { q^{m/2} - q^{-m/2} } \over {q^{1/2} - q^{-1/2}}}$)
$$ {\tilde {\bf L}}_{3i} \eq {\frac{1}{[2]_{\sqrt q} Y}} ~ \left[~~ 
\left({[-i~{\bf L}_{21}]_{{\sqrt q}}}\right)^2 ~,~
{\bf P}_{i} ~~\right] \plus {\bf P}_{i}~~, ~~~\left(Y \defeq \sqrt{{\bf P}^{1}{\bf P}^{1} ~+~ {\bf P}^{2}{\bf P}^{2}}\right) 
~~. \eqno(1.2)$$

Again, we have obtained the ``anti-deformation'' by solving eqns. (1.2) 
for the $ {\bf P}_{i}$.  Our result expresses the translation generators 
of  ${\cal E}(2)$ as irrational functions of $U_q(so(2,1))$ \cite{4}. If we let 
$ H = - ~2i~{\bf L}_{21}$, then our solution is given by: 
$${\bf P}_1 \eq D^{-1} ~\left(    
\{ 1 \minus { \frac {1} {2Y} } {\frac { [H]_q } {[H]_{\sqrt {q}} } } \}
{\bf L}_{31} \plus {\frac {i [2]_{\sqrt {q}} } {2 Y} } 
[ {\frac {H} {2} } ]_q {\bf L}_{32} \right) ~~, \eqno(1.3a)$$
and 
$${\bf P}_2 \eq D^{-1} ~\left(    
\{ 1 \minus { \frac {1} {2Y} } {\frac { [H]_q } {[H]_{\sqrt {q}} } } \}
{\bf L}_{32} \minus {\frac {i [2]_{\sqrt {q}} } {2 Y} } 
[ {\frac {H} {2} } ]_q {\bf L}_{31}\right) ~~, \eqno(1.3b)$$
where
$$ D \eq \minus {\frac {1 }{4 Y^2} } ~\left\{[ H ]_{\sqrt q}^2 \minus 
( {\frac { [H]_q } {[H]_{\sqrt {q}} } } \minus 2 Y)^2 \right\} ~~~. 
\eqno(1.4)$$
Furthermore
$$  Y^2 \eq {\tilde \Delta}_q  \plus \vier ~~~~ \eqno(1.5)$$
and ${\tilde \Delta}_q$ is the Casimir element of $U{_q}(so(3, \bbc))$ 
as defined in $[4]$.  
The reader can easily verify that the ${\bf P}_{i}$ as defined by eqns. (1.3a) 
and (1.3b) satisfy the defining commutation relations for the translation 
generators of  ${\cal E}(2)$, and 
verify that $Y^2$ satifies eqn. (1.5).

\section{{ ${ \bf S0_0(p+1,q), SO_0(p+1,q) \times_s \bbr^{p+q+1} }$ \bf and their Lie algebras}} 

Let $M$ denote the real vector space $ \bbr^{p+q+1}$ on 
which the quadratic form 
$$ Q(\xi) \eq \xi_0^2 +\xi_1^2 +~.~.~.~+\xi_p^2 -\xi_{p+1}^2 -~.~.~.~
-~\xi_{p+q}^2 ~~. \eqno(2.1)$$
is defined, 
and the associated bilinear form of signature $(p,q)$ is given by
$$B
(\xi,~\eta) ~=~ \halb \{Q(\xi+\eta)-Q(\xi)-Q(\eta)\}~. \eqno(2.2)$$
Denote the matrix 
corresponding to $B$ by $\b_0$: $\b_0 \eq { \rm diag} (1,~1,~.~.~1,~-1,~-1,~.~.~,~-1)$, where 
the right hand side of this equation 
denotes a diagonal matrix with diagonal entries as shown inside the 
parentheses, with $p+1$ entries being $+1$ and $q$  entries being 
$-1$.  

$SO_o(p+1,q)$ is the  component connected to identity of
 
$$SO(p+1,q) \eq  \{g \in SL(n,\bbr) ~\vert~ g\adj ~\b_0~g ~\eq ~\b_0 \}  ~~.\eqno(2.3)$$
($\adj$ denotes transpose of a matrix and $n \eq p+q+1$.) 
Its Lie algebra, $so(p+1,q)$, is defined as:

$$ {so(p+1,q)} \eq \{ X \in sl(n,\bbr) ~\vert ~ 
X\adj ~\b_0 \plus \b_0~ X ~=~ 0 \}~~.\eqno(2.4)$$

A basis of $\cg$ are the generators $ {\bL}_{ij}$ 
($i,~j \eq 0,~.~.~.~,~p+q$, $i~<~j$).  We define ${\bL}_{ij} \eq -{\bL}_{ji}$ 
for $i>j$, and the ${\bL}_{ij}$ satisfy the following commutation relations:
$$[{\bL}_{ij},~{\bL}_{jk}] \eq -e_{j}~{\bL}_{ik}  \eqno(2.5)$$
and all other commutators vanish.
The $e_j$ are defined as follows:
$$ e_{0} \eq e_{1} \eq 
~.~.~. \eq e_{p} \eq 1,~e_{p+1} \eq e_{p+2} \eq 
~.~.~. \eq e_{p+q} \eq -1  ~~. \eqno(2.6)  $$
Eqn. (2.5)   can also be given a well-known form 
with help of the $g_{ij}\eq (\b_0)_{ij}$:

$$[{\bL}_{ij},~{\bL}_{kl}] \eq  g_{ik}~  {\bL}_{jl} +g_{jl}~  {\bL}_{ik}-g_{il}~  {\bL}_{jk} ~- ~ g_{jk}~  {\bL}_{il} ~~, \eqno(2.7) $$

We wish to give explicit matrix 
expressions of these ${ {n(n~-~1)} \over 2}$ generators in our ($\b_0$) presentation of the fundamental representation 
of $\cg$. Let $E_{ij}$ denote the $n \times n$ matrix whose $i,j$ entry is 
one, and all other entries are zero.  Clearly
$$\left[E_{ij},E_{kl}\right] ~=~ \d_{jk}E_{il} ~-~ 
\d_{li}E_{kj}~~. \eqno(2.8)$$
Then ($q 
\ne 0$)
$${\bL}_{ij} \eq - \left(E_{i,j} ~-~E_{j,i} \right) ~~ (0 \le i,~ j \le p,~i \ne j) \eqno(2.9a)$$
$${\bL}_{n-i,n-j} \eq E_{n-i,n-j} ~-~E_{n-j,n-i} ~~ 
(1 \le i,~ j \le q,~i \ne j) \eqno(2.9b)$$
$${\bL}_{i,p+j} \eq E_{i,p+j} ~+~E_{p+j,i} ~~ 
(0 \le i \le p,~ 1 \le j \le q).  \eqno(2.9c)$$
The quadratic Casimir operator of $SO_0(p+1,q)$ is:
$${{\bQ}_2 } \eq \halb \sum_{i,j=0}^{p+q} ~~{\bL}_{ij}~{\bL}^{ji}~~\eq -~\halb ~  \sum_{i,j=0}^{p} ~~{\bL}_{ij}~{\bL}_{ij}~~
- $$
$$-~\halb ~  \sum_{i,j=p+1}^{p+q} ~~{\bL}_{ij}~{\bL}_{ij}~~+ ~~  
\sum_{i=0}^{p} ~~\sum_{j=1}^{q} ~~{\bL}_{i,p+j}~{\bL}_{i,p+j}~~. \eqno(2.10)$$

The Poincar\'e group in $p+q+1$ dimensions is the semi-direct product: 
$$SO_0(p+1,q) \times_s \bbr^{p+q+1}\eq \{ (g,a) \vert ~ g \in SO_0(p+1,q), ~~
a \in  \bbr^{p+q+1} ~~\}\eqno(2.11) $$
with multiplication law given by:
$$(g,a) ~~(g^\prime, a^\prime) \eq (g~g^\prime, ~ a ~+~ \phi(g)~a^\prime) 
\eqno(2.12)$$
where $\phi(g)$ is the natural action of 
$SO_0(p+1,q)$ on  $\bbr^{p+q+1}$.  
Its Lie algebra is 
$$ \cp \eq so(p+1,q) \oplus_s \bbr^{p+q+1},\eqno(2.13)$$
and a basis of $\cp$ is given by:
$$ {\bL}_{ij} ~~
~~, ~~{\bP}_k ~~(i,~j, ~k~ \eq 0,~.~.~.~,~p+q) ~~.\eqno(2.14) $$
The commutation relations of these basis elements are: eqns. (2.7) for 
the ${\bL}_{ij}$ and
$$[{\bL}_{ij}, {\bP}_k ] \eq  -~ g_{jk}~  {\bP}_{i} +g_{ik}~  {\bP}_{j}~~, ~~~~[{\bP}_{i}, {\bP}_j ] \eq 0 ~~.\eqno(2.15) $$
The following central element of the enveloping algebra of $\cp$ will 
play an important role in what follows: 

$${\bP}^2 \eq \sum_{k=0}^{p+q} {\bP}_k ~{\bP}^k \eq {\bP}_0^2 +{\bP}_1^2 +
~.~.~.~+{\bP}_{p+q}^2 -{\bP}_{p+q+1}^2 -~.~.~.~
-{\bP}_{p+q}^2  ~~.\eqno(2.16) $$

\section{Representations of 
${\bf SO_0(p+1,q)\times_s \bbr^{p+q+1} }$ from $ {\bf SO_0(p+2,q)}$ 
and ${ \bf SO_0(p+1,q+1)}$ representations}

It is well-know [1], [2] that the 
$$  {\bL}_{p+q+1,i}^\pm ~ = ~ {\frac {i~~[ {\bQ}_2, ~{\bP}_i] } {2~\sqrt{ \left(\pm ~ \sum_{i,~j~=~0}^{p+q+1} {\bP}_i ~{\bP}^{i} \right)}}} 
~+~{\bP}_i  \eqno(3.1\pm)$$
together with the ${\bL}_{i,j}$ of $SO_0(p+1,q)$ satisfy the defining 
commutation relations for the basic generators of $SO_0(p+2,q)$ or 
$SO_0(p+1,q+1)$ for the choice (3.1+) or (3.1$-$), respectively.
(Henceforth the plus sign of any $\pm$ refers to 
the $so(p+2,q)$ case ((3.1+)) and the minus sign of any $\pm$ refers to 
the $so(p+1,q+1)$ case ((3.1$-$)).) We can view eqns. (3.1$\pm$) as a system of equations for commuting 
quantities ${\bP}_i$, and 
we would like to solve these equations (3.1$\pm$) for the ${\bP}_i$. 
We have succeeded in doing this by 
working in a representation $(\pi, \ch)$ of 
$SO_0(p+1,q)\times_s \bbr^{p+q+1}$ (strongly continuous) on an 
Hilbert space $\ch$ for 
which:
$$ d\pi( ~{\bP}_{0}~{\bdel} ~~-
~~\sum_{i,j=1}^{p+q}~{\bP}_j~{\bL}_{oi}~{\bL}^{ij}) \eq 0  \eqno(3.2)$$
where
$$ {\bdel} \eq ~\halb~
\sum_{i,j=1}^{p+q} ~~{\bL}_{ij}~{\bL}^{ji}~~.$$
($d\pi$ denotes the representation 
of $\cp$ obtained from $\pi$ (by differentiation) and also its extension to 
$\ce(\cp)$, the enveloping 
algebra of $\cp$, and also to any algebraic extension of the Lie field of 
 $\cp$(c.f. ref. (1)).)  Now the quadratic Casimir 
operator of $so(p+2,q)$  $ \left[so(p+1,q+1) \right]$  constructed out of the 
${\bL}_{p+q+1,i}^+$ $ \left[{\bL}_{p+q+1,i}^- \right]$ of (3.1+) [(3.1$-$)] 
and ${\bL}_{ij}$ is: 
$${\bC}_{2}^\pm = {\bQ}_{2} - \sum_{i=0}^{p+q} {{\bL}^\pm}_{p+q+1,i}
{{\bL}^\pm}^{p+q+1,i} = {\bQ}_{2} \mp \sum_{i=0}^{p+q} {{\bL}^\pm}_{p+q+1,i}
{{{\bL}^\pm}_{p+q+1}}^{i}. \eqno(3.3\pm)$$

{\bf Lemma 3.1} For representations $(\pi, \ch)$ for 
which (3.2) holds the following is true: $d\pi({\bC}_{2}^\pm) \eq -~Y^2 
~-~\left( {{p+q} \over 2} \right)^2 ~\cdot 
~I\eq \mp d\pi({\bP}^2) ~-~\left( {{p+q} \over 2} \right)^2 ~\cdot 
~I~~,$ where $Y$ is the square root of the operator 
$\pm d\pi({\bP}^2)$, which we assume is self-adjoint, and strictly positive, 
so that it has a unique square root. Furthermore ($n=p+q+1$):
$$  d\pi({\bD}) ~ d\pi({\bP}_0) \eq  d\pi(\sum_{i=0}^{p+q} {{{\bA}_0}^i}~
{ {\bL}^\pm_{n,i} })~~\eqno(3.4\pm)$$
where
$${\bD}\eq { {(n-3)^2} \over 4 } \left\{{\bQ}_2 ~+~ 
{ {(n-1)(n-3)} \over 4 } \right\} ~+~i~Y~(n-3)~
 \{ {\bQ}_2 ~+~  $$ 
$$ 
{{(n-2)(n-3)} \over 4 } \}~-~Y^2~ \left\{ {\bQ}_2 ~-~ 
{ {(n-3)} \over 2 } \right\} ~+~i~Y^3~(n-2) ~-~Y^4,  $$
and
$$ \sum_{i=0}^{p+q} {{{\bA}_0}^i}~
{ {\bL}^\pm_{n,i} } \eq  {i \over 4} ~ (n-3)^2~ 
\left\{ ({ {n-3}  \over 2 }) ~ { {\bL}^\pm_{n,0} }  ~+~ 
\sum_{i=1}^{p+q}{ {{\bL}}^{0,i} } { {\bL}^\pm_{n,i} } \right\} ~Y $$ $$
 -~2 ~({ {n-3}  \over 2 }) ~ \left\{ \halb ~ ({ {n-3}  \over 2 }) ~ { {\bL}^\pm_{n,0} }  ~+~ 
\sum_{i=1}^{p+q}{ {{\bL}}^{0,i} } { {\bL}^\pm_{n,i} } \right\} ~Y^2  $$ $$~+ i~
\left\{ ({ {n-3}  \over 2 }) ~ { {\bL}^\pm_{n,0} }  ~-~ 
\sum_{i=1}^{p+q}{ {{\bL}}^{0,i} } { {\bL}^\pm_{n,i} } \right\} ~Y^3 ~ -~  { {\bL}^\pm_{n,0} } ~Y^4 ~~.$$
Now start with a unitary representation of $SO_0(p+2,q)$  
$[ SO_0(p+1,q+1)]$ on an Hilbert space $\ch$ and 
we use eqn. (3.4$\pm$) to 
construct a representation of $\cp$ out this representation. 
Denote the (abstract) generators of  $SO_0(p+2,q)$  
${ \left[ SO_0(p+1,q+1)  \right]}$ by ${\tbL}_{ij}$ ($i=0,1,...,p+q$) 
and ${\tbL}^{+}_{n,i}$ ${ [ {\tbL}^{-}_{n,i} ]}$ 
($n=p+q+1$, $i=0,1,...,p+q$).

{\bf Theorem 3.1}  Let $(\pi^\pm, \ch)$ be a continuous, unitary representation of $SO_0(p+2,q)$ [$SO_0(p+1,q+1)]$
on an Hilbert space $\ch$ and let  $(\pi^\pm, \ch)$ be such that:
$$ \left( \psi, \left\{ d\pi^\pm( {\tbC}_2) +  \left( { {p+q}  
\over 2 } \right)^2 \cdot  
I \right\} ~\psi \right) \gl 0 ~\forall~  \psi ~\in~ {\cd}(d\pi( {\tbC}_2))~, 
\eqno(3.5\pm) $$
where ${\cd}(~~)$ denotes the domain of an operator ($>$ is for (3.5+) 
and $<$ is for (3.5$-$)) and 
$$ d\pi^\pm \left( ~{\tbL}^{\pm}_{n0}~{\tbdel} ~~-
~~\sum_{i,j=1}^{p+q}~{\tbL}^{\pm}_{nj}~{\tbL}_{oi}~{\tbL}^{ij}\right) \eq 0 
~~. \eqno(3.6)$$
Let $Y:\ch
\longrightarrow ~\ch$ be a bounded, symmetric linear operator on $\ch$, which commutes 
with all elements of $d\pi^+(so(p+2,q))$ [$d\pi^-(so(p+1,q+1))$] 
and satisfies 
$$ Y^2 \eq - \left\{ d\pi({\tbC}_2) ~+~   \left({ {p+q}  \over 2 }\right)^{2}~\cdot ~ 
I \right\} ~~.\eqno(3.7\pm)$$
Further, let ${\tbD}$ and  $\sum_{i=0}^{p+q} {{{\tbA}_0}^i}~
{ {\tbL}^{\pm}_{n,i} }$ be as in eqn. (3.4$\pm$), but with  ${\tbL}_{ij}$ and ${\tbL}^{\pm}_{n,i}$ instead 
of ${\bL}_{ij}$ and ${\bL}^{\pm}_{n,i}$. Then there is formally a skew 
symmetric 
representation $d\tpi^\pm(\cp)$ of 
$\cp = so(p+1,q) ~ \oplus_s ~ \bbr^{p+q+1}$ 
on $\ch$ with $d\tpi^\pm( {\bL}_{ij}) =  
d\pi^\pm( {\tbL}_{ij})$
$(i = 0,1, ...,p+q)$, $d\tpi^\pm( {\bP}_{0})= d\pi^\pm({\tbD})^{-1} ~
d\pi^\pm(\sum_{i=0}^{p+q} {{{\tbA}_0}^i}~{ {\tbL}^{\pm}_{n,i} })$, 
and 
$d{\tpi^\pm}( {\bP}_{i}) = [d\tpi^\pm( {\bL}_{i0})$, 
$d{\tpi^\pm}( {\bP}_{0}) ] (i= 1, 2, ..., p+q)$, 
provided  $d{\pi}^{\pm} \left( \left[{\tbQ_2} , 
\sum_{i=0}^{p+q} {{{\tbA}_0}^i}~
{ {\tbL}^{\pm}_{n,i} }  \right] \right) \ne 0$, $d\pi^\pm({\tbD})^{-1}$ exists on a suitable dense domain in $\ch$, and the operators 
$d\pi^\pm({\tbD})^{-1}~d{\pi} \left( \left[{\tbL}_{i0} , \sum_{j=0}^{p+q} {{{\tbA}_0}^j}~
{ {\tbL}^{\pm}_{n,j} }  \right] \right)$ ($i ~=~ 1,~2,~3$) and 
$ d\pi^\pm({\tbD})^{-1} ~
d\pi^\pm(\sum_{i=0}^{p+q} {{{\tbA}_0}^i}~{ {\tbL}^{\pm}_{n,i} })$ commute with 
each other.

\section{The  General Solution of the Problem in the 
${\bf SO_0(3,2)}$ case and Tachyons } 

Now we consider the general solution of eqns. (3.1+) 
for $p=0$ and $q=3$, without imposing any conditions, such 
as those of eqns. (3.2) and (3.6).  In order to describe the solution 
it is necessary to know a bit more about the structure of the enveloping 
algebras $\ce(\cg)$ of $\cg = so(2,3)$ and $\ce(\cp)$, where $\cp$ is 
the Poincar\'e Lie algebra.

A maximal abelian subalgebra of  $\ce (\cg)$ is generated by the
six operators:
$${\bL}_{12}~,~{\bL}_{12}^{2} ~+~ {\bL}_{23}^{2} + {\bL}_{31}^{2} ~~=~~
{\bL}^{2}~,{\bQ}_{2} ~~=~~ {\bL}_{01}^{2} ~+~ {\bL}_{02}^{2} ~+~
{\bL}_{03}^{2} ~-~ {\bL}^{2} $$
$$~ {\bQ}_{4} ~~=~~ \left( {\bL}_{12} ~{\bL}_{30} ~+~ {\bL}_{23} ~{\bL}_{10} ~+~ {\bL}_{31} ~{\bL}_{20}\right)^{2} ~,~  
{\bC}_{2} ~,~ {\bC}_{4} \eqno(4.1)$$ 
where
$$ {\bC}_{2} ~=~   -{\bL}_{04}^{2} +{\bL}_{01}^{2} +{\bL}_{02}^{2} 
+{\bL}_{03}^{2} - {\bL}_{12}^{2} - {\bL}_{23}^{2} - {\bL}_{31}^{2}
+ {\bL}_{14}^{2} + {\bL}_{24}^{2} + {\bL}_{34}^{2} \eqno(4.2)
$$
and
$$ {\bC}_{4} ~=~ -{\left[ {1 \over 4} ( \bl ~-~ \br ) \right]}^{2}
~-~ \left( {\bL}_{12} ~{\bL}_{30} ~+~ {\bL}_{23} ~{\bL}_{10} ~+~ {\bL}_{31} 
~{\bL}_{20}\right)^{2}   ~+ $$
$$+ \left( \sum^3_{{i~j~k  \atop  \ell~ m}~=~1} ~(\e_{i~j~k} ~\{ {1 \over 2} 
{\bL}_{04} {\bL}_{jk} -
{\bL}_{0j} {\bL}_{4k} \})(\e_{i~\ell~m} \{ {1 \over 2} {\bL}_{04} 
{\bL}_{\ell m} -
{\bL}_{0 \ell} {\bL}_{4m} \}) \right) . \eqno(4.3) $$
with 
${1 \over 4} ~( \bl ~-~ \br ) ~~=~~ {\bL}_{12}~ {\bL}_{34} ~+~ {\bL}_{23}~ {\bL}_{14}
~+~ {\bL}_{31} {\bL}_{24}$. $ - {\bL}^{2}$ and $-i {\bL}_{12}$ are 
operators for the total angular 
momentum squared and the third component of the total angular momentum,
respectively.  The center $ Z  ( {\cg}  ) $ of the universal 
enveloping algebra 
$\ce (\cg)$   of $SO_0(2,3)$ is generated by ${\bC}_2$ and ${\bC}_4$.

The center ${ Z}(\cp)$ of the enveloping algebra of $\cp$ 
is generated by the following set of elements: 
$${\bP}^2 \eq \sum_{k=0}^{3} {\bP}_k ~{\bP}^k \eq {\bP}_0^2 -{\bP}_1^2 
~-~{\bP}_{2}^2 -{\bP}_{3}^2 ~~,\eqno(4.4) $$
and 
 $${\bw} ~=~ \sum_{\mu=0}^{3} \sum_{\nu=0}^{3} \sum_{\rho=0}^{3} \left(
{\bP}_\mu ~ {\bP}^\nu ~{\bL}_{ \nu \r } ~ 
{\bL}^{ \r \mu} ~-~ 
{ 1 \over 2} {\bP}_\r ~ {\bP}^\r ~{\bL}_{ \mu \nu } ~ {\bL}^{ 
\nu \mu} \right) ~. \eqno(4.5)$$
$ -~{\bP}_\mu ~{\bP}^\mu$ is the operator for the square of the mass, 
and ${\bw}$ is 
a scalar operator, which describes the spin in a relativistically 
invariant way. 

Now let ${\bC}^{\prime}_{2}$ 
$=$ 
$-\left({ \bC }{_2} + \5in2  ~ \bf I \rm \right)$ 
and ${ \bC}^{\prime}_{4}$ 
$=$ 
$~-~({ \bC }{_4}$ 
$-$ 
$ \vier $ 
${ \bC }{_2} $ 
$-$ ${ 9 \over 16} ~ \bf I \rm )$. ($ \bf I \rm $ is the identity 
in $\ce (\cg)$.)  We have the following result: 
 
{\bf Theorem 4.1}  

$${\bP}_{\mu} ~=~ { {\bD}}^{- 1} ~
{{\bA}^{~~\nu}_{\mu}}
 {\bL}_{\nu  \rm 4}  \eqno(4.6)$$
with 
$${\bf  A}^{~~\nu}_{\mu} ~=~ - { \bf  C }^{\prime}_{4} ~\d_{\mu}^{~~\nu} 
~+~ {\frac {i} {2}} \left[ 
\left( {\bQ}_{2} ~+~ \vier \right) \d_{\mu}^{~~\nu} ~-~ 
{\threehalves} {\bL}_{\mu}^{~~\nu} 
~-~ {\bL}_{\mu  \r } {\bL}^{ \r \nu} ~-~ {\bQ}_{4} \e_{\mu ~~\r \t}^{~~\nu} 
{\bL}^{\r \t} 
\right] ~ {\bY}  $$
$$~-~ \left[ \left({\bQ}_{2} ~+~ \vier ~-~  
{ \bC}^{\prime}_{2} \right) \d^{~~\nu}_{\mu} ~-~ 
{\bL}^{~~\nu}_{\mu} ~-~ 
{\bL}_{\mu \r} {\bL}^{\r \nu} \right] { {\bY}^{2} } 
~+~i~ \left( \halb \d^{~~\nu} _{\mu}  ~-~
{\bL}^{~~\nu} _{\mu} \right) { {{\bY}^{3}}  }   
\eqno(4.7)$$
and 
$$ {\bD} ~~=~~ \left( {\bQ}_{4} ~+~ \vier {\bQ}_{2} ~-~  
{ \bC}^{\prime}_{4} ~+~ { 3 \over 16} ~{\bf I} \right) ~+ $$ $$~+~ 
{i} \left( {\bQ}_{2} 
~+~ {\frac{1}{2}} \right) 
 {{\bY} } ~-~ \left( {\bQ}_{2} ~-~
  { \bC}^{\prime}_{2} ~-~ 
\halb \right) { {{\bY}^{2}}  } ~+~ { {2 ~i~ {\bY}^{3} }} ~. 
\eqno(4.8)$$  
Furthermore ${\bY}^2$, which commutes with all elements of $\ce (\cg)$, satisfies the following equation
$$ {\bY}^4 ~+~  {\bC}^{\prime}_{2} ~ {\bY}^2  +    { \bC}^{\prime}_{4} 
\eq 0 ~~. \eqno(4.9) $$

The reader may readily convince himself that in spin zero case, 
for which ${ \bC }{_4}$ and 
${ \bQ }{_4}$ are represented as the zero 
operator, eqns. (4.6) (with eqns. (4.7) and (4.8)) for ${\bP}_{0}$ (in a 
given representation which satisfies eqn. (3.6)), agree 
with the corresponding 
equations (3.4+) in the Lemma 3.1. when $p=0$, $q=3$ . The main fact, which 
is necessary in order to 
convince oneself of this, is the following: 
$$d\pi^+\left({\tbL}_{0\rho}~{\tbL}^{\rho \nu}~{\tbL}^{+}_{4\nu}\right) \eq 
d\pi^+\left({\tbQ}_2 {\tbL}^{+}_{40} ~-~ 2~{{\tbL}_{0}^{~~i}}~{\tbL}^{+}_{4i}
\right)
~~,$$
which is easily proved with the help of the representation condition, 
eqn. (3.6).

Note that ${ {\bD}}^{- 1}$ makes sense because $\ce (\cg)$ has no 
zero divisors \cite{5}. Also eqn. (4.9) is much more interesting  than the 
corresponding equation in 
the spinless case of the previous section, which we obtain out of 
eqn. (4.9) by setting ${ \bC}^{\prime}_{4}$ = 0. Eqn. (4.9)  is a quartic equation 
for the operator $\bY$ involving both Casimir operators of 
$SO_0(2,3)$, and thus there are many more possibilities with which 
to contend, e.g. 
de Sitter mass becoming Poincar\'e spin and de Sitter spin becoming Poincar\'e 
mass. 

Finally, we describe a class of spinless representations of $SO_0(2,3)$ 
which satisfy the hypotheses of Theorem 3.1. We show that the representations of the Poincar\'e Lie algebra which we get from this theorem correspond 
to infinitesmally unitary representations with imaginary mass, i.e. they 
are tachyonic representations. (It seems clear to us that they should be 
integrable to Poincar\'e group representations, although we do not 
have a rigorous proof of this fact.) 

The $SO_0(2,3)$ representations 
which we consider are continuous series representations occuring in 
the left regular representation on anti-de Sitter space. These representations and those of the continuous series occuring in the decomposition of 
the left regular representation of $SO_0(2,3)$ on anti-de Sitter space 
into irreducibles is a special case of a very 
beautiful achievement of 20th century mathematics,
namely: the decomposition into 
irreducibles of the left regular representation of $SO_0(p+1,q)$ on real hyperbolic 
spaces. Let us briefly describe this result here. 

Real hyperbolic spaces $H^{p,q}$ ($q>0$) are: 
$$ H^{p,q} =
\{ \xi \in \bbr^{p+q+1} \vert \xi_0^2 +\xi_1^2 +\xi_2^2 +
...+\xi_p^2 -\xi_{p+1}^2 -...-\xi_{p+q}^2  \eq -1 \}~~.\eqno(4.10)$$  
The isotropy subgroup of the point $ \xi^i = 0$ ($i \ne p+q$), $\xi^{p+q}= 1$ is 
$H=SO_0(p+1,q-1)$. Since the action of $SO_0(p+1,q)$ is transitive on $H^{p,q}$,we have that 
$$H^{p,q} \simeq SO_0(p+1,q)/SO_0(p+1,q-1) ~~.\eqno(4.11)$$ 
$H^{p,q}$ are semi-Riemannian spaces of constant curvature.  The 
representation of $G=SO_0(p+1,q)$ is the left regular representation on 
$H^{p,q}$ i.e. $(\pi(g)f)(\xi)=f(g^{-1}~\xi)$ 
for any $g \in G$ and $f \in \cl^2(G/H)$. We have on $C^\infty(G/H)$ 
representations $d\pi(so(p+1,q))$ and $d\pi(\ce(so(p+1,q))$ of the 
Lie algebra and the enveloping algebra, respectively.
Let 
$$\ch_s = \{ f \in  C^\infty(G/H)  \vert   
 (d\pi({\bC}_2)f) = \left( s^2 ~-~  \left\{ { {p+q-1} \over {2} } 
\right\}^2 \right) f \}\eqno(4.12)$$ 
for each $s \in \bbc$.  We then have: \cite{6}

{\bf Proposition 4.1} $d\pi({\bC}_2)$ has a self-adjoint closure in 
$\cl^2(G/H)$. For $q>1$ the spectrum of $d\pi({\bC}_2)+[(p+q-1)/2]^2 
\cdot I$ in 
$(0,~ \infty)$ is discrete with eigenvalue $[s + (p+q-1)/2]^2$ for 
each integer $s > -(p+q-1)/2$.  Corresponding to each eigenvalue 
is an infinite dimensional eigenspace $\ch_s$.  The completion of 
$\ch_s$ in the $\cl^2(G/H)$ norm gives an irreducible, unitary 
representation of  $SO_0(p+1,q)$. The spectrum of 
$d\pi({\bC}_2)+[(p+q-1)/2]^2 \cdot I$ in $(- \infty, 0]$ is continuous 
with eigenvalue $[s]^2$ for 
each  $s \in i~\bbr$.  The completion of 
$\ch_s$ in the $\cl^2(G/H)$ norm is the direct sum of two irreducible, 
unitary representations of  $SO_0(p+1,q)$. For $q=1$ the above is also 
true except that there is no discrete spectrum.

We leave it to the reader to convince himself that the representations of 
this Proposition fulfill the hypotheses of Theorem 3.1. (Since 
$d\pi({\bC}_2)$ and $d\pi({\bQ}_2)$ are self-adjoint it is clear that $d\pi^+({\tbD})$ is 
invertible. To prove the mutual commutativity of the translation generators 
one uses an integral transform given in \cite{7}.)  The representations of the continuous 
spectrum, are the ones we use with (3.5+) ($p$ of 
proposition replaced by $p+1$). The representations of 
the continuous spectrum are also the ones we use with (3.5$-$) ($q$ in proposition replaced by 
$q+1$). For the  $SO_0(2,3)$ case ((3.5+)) we see that  
$d{\tpi^+}( {\bP}^2)$ is 
strictly greater than zero, and thus these representations are 
tachyonic \cite{8} as claimed above. We are also considering application of 
these ideas to higher dimensional cases \cite{9}~.
\begin{acknowledgements}
The author wishes to thank Professors M. Havlicek for useful 
discussions; it is also a pleasure to thank Professor H.P. Jakobsen for 
some valuable comments.
\end{acknowledgements}

\end{document}